\documentclass[twocolumn]{aastex62}

\received{5/18/2018}
\revised{7/11/2018}
\accepted{7/24/2018}
\submitjournal{ApJ}

\shorttitle{MHO 3252 Y3 Variability}
\shortauthors{Hodapp,Chini}

\begin{document}
\title{Variability and Jet Activity in the YSO MHO~3252~Y3
in Serpens South}

\correspondingauthor{Klaus Hodapp}
\email{hodapp@ifa.hawaii.edu}

\author[0000-0003-0786-2140]{Klaus W. Hodapp}
\affil{University of Hawaii, Institute for Astronomy \\
640 N. Aohoku Place \\
Hilo, HI 96720, USA}

\author{Rolf Chini}
\affiliation{
Astronomisches Institut, Ruhr-Universit{\"a}t Bochum \\
Universit{\"a}tsstra{\ss}e 150 \\
44801 Bochum, Germany
}
\affiliation{
Instituto de Astronomia, Universidad Catolica del Norte \\
Avenida Angamos 0610 \\
Antofagasta, Chile
}



\begin{abstract}

The infrared young stellar outflow source
MHO~3252~Y3 in the Serpens South star-forming region
was found to be variable. 
The available photometric data can be fitted with
a double peaked light curve of 904~d period.
Color variations are consistent with
variable extinction with a flatter wavelength dependence than
interstellar extinction, i.e., larger grains.
MHO~3252~Y3 is the source of a large scale bipolar outflow,
but the most recent outflow activity has produced a microjet
detectable in the shock excited H$_2$ 1--0 S(1) line while
[\ion{Fe}{2}] emission appears confined to the immediate vicinity
of the central star.
The proper motion of individual shock fronts in the H$_2$ microjet has been measured
and traces these knots back to ejection events in the past
two centuries.
Integral field spectroscopy with the Keck~1 adaptive optics system
and the OSIRIS instrument shows velocity components
near the launch region that are distinct from the
microjet both in radial velocity and apparent proper motion.
They match the prediction of dual wind models with a
distinct low velocity disk wind component. We find evidence for 
the entrainment of this low velocity component into the high velocity
microjet, leading to shock excited emission at intermediate velocities
in an envelope around the microjet.

\end{abstract}




\section{Introduction} \label{sec:intro}

The accretion process forming stars out of much larger molecular clouds
can only proceed if the excess angular momentum of the original -- usually rotating --
molecular cloud material can be disposed of. This is achieved by the 
ejection of a fraction of the infalling material in the form of bipolar
molecular outflows. The first such outflow was discovered by \citet{Snell.1980ApJ...239L..17S}
in millimeter CO emission. 
Faster, more collimated jets of molecular and atomic material were discovered
at optical wavelengths by proper motion studies of HH objects
\citep{Herbig.1981AJ.....86.1232H} and by the first imaging studies of the jets
themselves
by \citet{Dopita.1982ApJ...263L..73D}, \citet{Graham.1983ApJ...272..615G}, 
\citet{Mundt.1983ApJ...265L..71M}, and
\citet{Mundt..1983ApJ...274L..83M}.
The region where jets are launched and then collimated
is so close to the central star that observations with the highest achievable spatial resolution
are required.
Velocity resolved observations of the first few hundred
astronomical units of a jet have been done with the Hubble Space Telescope (HST) at optical and
near-infrared wavelengths, and most recently with integral field spectrographs
and adaptive optics systems on ground-based large telescopes.
However, only a small number of protostellar jet sources
could, so far, be studied
because they either
must be optically detectable for HST or have an optically visible nearby tip-tilt reference star
for ground-based adaptive optics work in the near infrared.

This paper does not intend to present a review of the extensive literature
on protostellar outflows and jets. The reader is referred to 
\citet{Reipurth.2001ARA&A..39..403R} for a review of the older literature,
and the more recent review 
by \citet{Frank..2014..prpl.conf..451F} at the Protostars \& Planets VI Conference.

While all the cases studied in detail show large-scale outflow, a diversity of phenomena has emerged
near the driving source, probably dependent on the mass, evolutionary
state, and orientation of the object.
Among the jet sources studied with sufficient detail
both classical jet features as well as series of bubbles have been observed.
In some cases, microjets of only a few arcseconds in length can be observed 
that kinematically are only decades to a few centuries old and therefore
offer the best opportunity for studying the conditions of their launch.
Some objects actually show both
a jet and bubbles in different excitation tracers.
The very young outflow phase T~Tauri star XZ~Tau was imaged repeatedly by
\citet{Krist2008} using HST and shows a series of bubbles rather than a typical jet
while the immediately adjacent
HL~Tau shows classical jet features.
A similar pattern of bubbles traced in the H$_2$~1--0~S(1) line has been observed
in NGC~1333~SVS13
studied by \citet{Hodapp2014ApJ...794..169H}. In this outflow object a jet is observed in the
high-excitation [\ion{Fe}{2}] line, while the lower excitation H$_2$~1--0~S(1) line
shows the morphology of a string of bubbles.
Another example of spherically expanding shock fronts, rather
than jets, is the high-mass protostar in Cepheus~A~HW2, where
\citet{Torrelles.2001Natur.411..277T} 
observed an expanding ring of H$_2$O masers.

On the other hand, objects showing a well collimated microjet
generally exhibit
a complex velocity structure,
as was reviewed in, e.g., 
\citet{Frank..2014..prpl.conf..451F}: 
A high velocity component (HVC) associated with the
well collimated microjet itself, and a low velocity component that is
spatially distinct, broader and best observable in mm-wave emission lines of CO. 
We will, in the following, refer to the latter as the broad low velocity component (BLVC).
Shock-excited emission from such a BLVC is usually identified with the interaction of
a low velocity disk wind with the outflow cavity.
Detailed velocity studies of YSO jets often show velocity gradients in the jet
from the HVC on the jet axis to an intermediate velocity component (IVC) 
forming an envelope around the jet.
Its formation and excitation is best explained by lateral entrainment
of the BLVC wind by the HVC of the jet, as discussed in detail for the case of
DG Tau by
\citet{2016MNRAS.455.2042W}, even though this issue is still actively being debated.

We will frequently refer to DG Tau as a comparison case in the discussion
of our object MHO~3252~Y3, and therefore introduce DG Tau in some level of detail.
The DG Tau jet was first imaged by
\citet{Mundt..1983ApJ...274L..83M} 
and then studied spectroscopically
in forbidden emission lines at optical wavelengths by
\citet{Bacciotti.2002ApJ...576..222B}, and in 
[\ion{Fe}{2}] by
\citet{Pyo.2003ApJ...590..340P} with Subaru IRCS.
A detailed study of the proper motion of the DG~Tau jet knots
based on VLA radio, optical, and X-ray
data was presented by
\citet{Rodriguez.2012A&A...537A.123R}.
They found that optical/infrared emission knots coincide with the VLA radio knots,
identify one X-ray emission knot earlier seen in optical emission and present
a geometric model of the DG~Tau bipolar outflow.

Velocity-resolved integral field spectroscopy was presented by \citet{Agra-Amboage.2011A&A...532A..59A} and\\
\citet{Agra-Amboage.2014A&A...564A..11A} 
with the ESO/SINFONI integral field spectrograph
and independently by
\citet{White.2014MNRAS.442...28W} and
\citet{2016MNRAS.455.2042W} using Gemini/NIFS.
The microjet in DG~Tau is detected in
[\ion{Fe}{2}] emission, while the H$_2$ 1--0 S(1) emission line
shows only a BLVC.
The  blueshifted part of the DG~Tau microjet has two velocity components in
[\ion{Fe}{2}]:
A well collimated high velocity (HVC) microjet, enveloped by an intermediate velocity 
component (IVC). 
The redshifted outflow lobe of DG~Tau, however, shows only the wider
BLVC with a bubble-like structure traced both in the
[\ion{Fe}{2}] and
H$_2$ 1--0 S(1) emission lines.

The DG~Tau microjet is comprised of multiple emission knots, formed by internal
shocks within the jet. Their spacing indicates that events leading to the
formation of different velocity ejecta must occur every few years.
Older determinations of the bubble spacing by
\citet{Pyo.2003ApJ...590..340P}
indicated a 5~yr period, confirmed by
\citet{Rodriguez.2012A&A...537A.123R}
while 
\citet{Agra-Amboage.2011A&A...532A..59A} claimed that the ejection period
is even shorter at 2.5~yr.
\citet{2014MNRAS.441.1681W} 
discussed all available studies of the ejection interval and in the end
concluded that the ejection interval is
variable and subject to future studies.

Any accretion process from a rotating molecular cloud onto a protostar
must solve the problem of excess angular momentum of the infalling gas.
Rotation of the jet would provide a mechanism for ejecting angular momentum and
for allowing most of the infalling matter to accrete onto the star.
Observational confirmation that protostellar and T~Tauri jets do indeed
rotate has proven exceedingly difficult, however. The state of this field has recently been
reviewed by 
\citet{Frank..2014..prpl.conf..451F}. 
In their paper on the ultimately unsuccessful search for rotation in the RY Tau jet, \citet{Coffey.2015ApJ...804....2C}
noted that for a convincing detection of rotation, it should be detected near the driving
star, should be consistent along the jet, should rotate consistent with the rotation
of the disk, and be independently confirmed by other observers. So far, no optical/NIR jet
has satisfied this strict set of criteria. Even for the best studied jet at optical
and near-infrared wavelengths, DG~Tau, which we introduced above, rotation studies in different
emission lines and by different authors remain inconsistent.
 
There have been some successful detections of rotational signatures at sub-mm and mm
wavelengths, using the superior velocity resolution of heterodyne radio receivers, e.g.,
CB~26 \citep{Launhardt.2009A&A...494..147L},
DG~Tau~B \citep{Zapata.2015ApJ...798..131Z},
Ori-S6 \citep{Zapata.2010A&A...510A...2Z},
NGC~1333~IRAS~4A2 \citep{Choi.2011ApJ...728L..34C},
and HH~797 \citep{Pech.2012ApJ...751...78P},
while others ascribed observed asymmetric kinematic features of jets to the effects of orbital
motion, e.g. 
\citep{Lee.2009ApJ...699.1584L} and
\citep{Lee.2010ApJ...713..731L}.
A particularly convincing example of rotation in the protostellar jet HH~212 has recently
been published by
\citet{2017NatAs...1E.152L} 
and 
\citet{2018ApJ...856...14L} using ALMA data of high spatial and spectral resolution.

The Serpens South star forming region, discovered by \citet{Gutermuth-2008ApJ...673L.151G} 
is the site of multiple very young stars
in their accretion and outflow phases. 
The most reliable distance to the Serpens complex of molecular clouds, of which
Serpens~S is a part, has been measured by \citet{Ortiz-Leon2017} to be 436.0$\pm$9.2~pc,
based on VLA parallax measurements of several radio-bright YSOs in that cloud.
We adopt this distance here
and refer the reader to the
extensive discussion in the paper
by \citet{Ortiz-Leon2017}
of different distance determination methods, older distance measurement,
the spatial structure of the Serpens molecular
cloud, and its association with larger structures.
In the Appendix, we also briefly discuss one star near Serpens South with a {\it Gaia}
\citep{Gaia-2016A&A...595A...1G} 
DR2 distance
measurement and conclude that its measured distance is consistent with that 
determined by \citet{Ortiz-Leon2017}.

\citet{Teixeira2012} have conducted a search for molecular
hydrogen shock fronts of bipolar outflows of young stellar objects (YSO) 
in the Serpens South region, and have included
many of them in the catalog of Molecular Hydrogen Objects (MHO) \citep{Davis..2010A&A...511A..24D} 
\footnote[1] {The MHO catalog is now hosted by D. Froebrich at the University of Kent, U.K.}.
In particular, the outflow source listed as MHO~3252~Y3 by them
at $\rm 18^h30^m01\fs3~-2\arcdeg10\arcmin26\farcs0~(J2000)$ 
has attracted our attention as a variable infrared
object with substantial amplitude and a 904~d~period.
These observations will be described in section 2.1 and 3.1.
MHO~3252~Y3 is the driving source of a short,
highly collimated microjet with multiple shock fronts, as well as a more extended
system of older shock fronts, traced in the shock-excited H$_2$~1--0~S(1) line.
As an overview of this object we show an RGB false color composite of UKIRT WFCAM $H$, S(1) line, and $K$-band
images in Figure~1. The $S(1)$ line image (green) is substantially deeper than the
image by
\citet{Teixeira2012} and shows a well developed, though faint bow shock
in the eastern outflow lobe. Proper motion measurements, discussed later in
chapter 3.2, are also indicated in Figure~1.

\begin{figure*}
\begin{center}
\includegraphics[angle=0,scale=1.0]{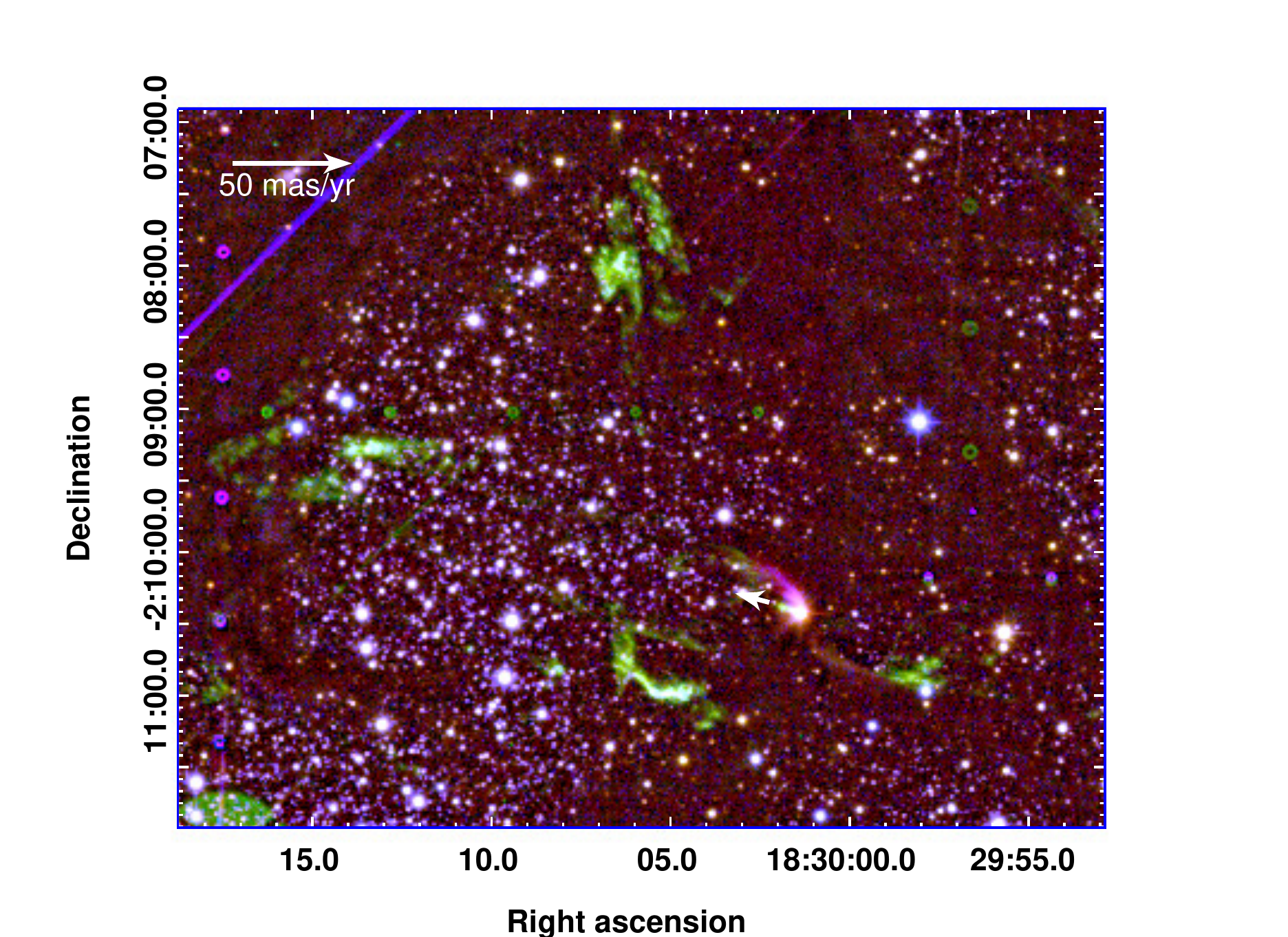}
\caption{
Three color composite of $H$-band (blue), $S(1)$ line emission (green), and $K$-band (red)
UKIRT WFCAM images of the Serpens South MHO~3252~Y3 region, obtained in 2016 and 2017. Shock-excited features show up in green. MHO3252~Y3 is the source
of a large-scale well developed bipolar outflow.
Proper motion vectors based on the 2011 Calar Alto data of \citet{Teixeira2012} and on 2016 UKIRT data are indicated.
}
\end{center}
\end{figure*}

The small extent of the microjet of only a few arcseconds and the availability of a sufficiently
bright star for tip-tilt correction make this object a suitable
target for adaptive optics integral field spectroscopy with the OSIRIS
instrument on Keck~1, allowing a combination of diffraction-limited spatial
and medium spectral resolution.
These observations will be presented in sections 2.2, 3.2, 3.3, and 3.4
and discussed in section 4.

There are only a few examples of microjets of young low mass stars that
are suitable for detailed studies. Each of these examples show individual
peculiarities, and a generalized picture of jet formation can only
be formed by studying as many as possible of the available cases.
This paper adds MHO~3252~Y3, a small, relatively low excitation microjet, to the list
of well studied cases of protostellar jets.

\section{Observations and Data Reduction}

\subsection{IRIS and UKIRT Imaging Photometry}

As part of an on-going program of monitoring star-forming regions for variable objects,
from 2012 on,
the area of MHO~3252~Y3 was repeatedly imaged
in the $K_s$ band
\citep{Skrutskie2006}
with the IRIS 0.8m telescope
\citep{Hodapp2010}
and 1024$\times$1024 2.5~$\mu$m IRIS infrared camera 
of the Universit\"atssternwarte Bochum
on Cerro Armazones, Chile.
This infrared camera was described in detail under its old name (QUIRC) and in
its original optical configuration by 
\citet{Hodapp1996QUIRC}.
For each visit of a specific region, IRIS takes 20 raw data frames on the object with an
integration time of 20~s, with small dithers
to get around detector defects, followed by 20 dithered data frames on a separate sky field
chosen to avoid bright stars or extended nebulosity. For these sky exposures, larger dithers
are used than for the object exposures to minimize the impact of any nebular emission on
the computation of the sky frame.

The source MHO~3252~Y3 reaches $K_s\approx$~10 in maximum brightness, which is near the bright limit
for the standard 20~s exposures with IRIS used for monitoring. We therefore base the light curve
of this object on the "short" IRIS exposures. After detector reset and the first of the reads
of the double correlated pair of readouts, the IRIS camera immediately reads the detector out a second
time, before the rest of the 20~s nominal integration time elapses. The difference of these
two frames early in the integration time is a double-correlated pair with an effective integration
time of 2~s. The sensitivity of this short
integration is entirely adequate even for recording the $K_s\approx$~12 minimum of the MHO~3252~Y3 light curve
and therefore the IRIS photometry of MHO~3252~Y3 is based on these short exposures.

The IRIS camera raw data were processed in a reduction pipeline written by R. Watermann and based on
the Image Reduction and Analysis Facility (IRAF) software \citep{Tody1986}.
The raw data were flatfielded using incandescent light differential dome flats.
Sky frames are computed as the sky-background adjusted median of the sky position frames taken
at some distance from the star-forming region and its extended nebular emission,
and then subtracted from the object frame.
The optical distortion correction was based on an as-built ZEMAX model of the optical
system. After correcting the known optical distortions, the
astrometric solutions of the individual images were calculated using SExtractor \citep{Bertin1996} and
SCamp \citep{Bertin2005} and based on Two Micron All Sky Survey (2MASS) coordinates \citep{Skrutskie2006}.
The images were then co-added using these astrometric solutions.
Aperture photometry was obtained with the IRAF task PHOT with an aperture of $r = 4$~pixels ($r = 3\arcsec$)
and calibrated against an ensemble of stars from the 2MASS
catalog. These 2MASS reference stars were selected for appropriate brightness,
and for being isolated from other stars and from nebular emission.
The 2MASS reference stars were in the brightness range K$_s$ = 12.5 - 14.5.
For part of the monitoring period, the IRIS telescope experienced unusually large pointing errors and
unstable image quality due to a mirror support problem, so that
there are only 9 suitable 2MASS photometric reference stars common to all IRIS frames used here.
The rms scatter of the measured magnitudes ranged from
0.03 for the brightest to 0.10 mag for the faintest reference stars.

In addition to the monitoring with IRIS, deeper images with better spatial resolution
were obtained with the Wide-Field Camera (WFCAM) \citep{Casali2007} on UKIRT in 2016 and 2017.
The WFCAM filters conform to the ''MKO'' standard described by
\citet{Tokunaga2002} and further characterized in
\citet{Tokunaga2005PASP..117..421T}.
As is done in the UKIRT manuals, we will refer to the UKIRT
$K$ bandpass simply as ''$K$''.
The UKIRT images in the $J$, $H$ and $K$ broad bands and in the narrow S(1) line filter
are substantially deeper than those of \citet{Teixeira2012} and show
additional details of the shock fronts.
The UKIRT WFCAM data were processed by the pipeline run by Cambridge Astronomy Survey Unit (CASU),
and the delivered FITS image files were directly used for photometry.
For the purpose of determining the light curve, we did not do any color
correction between the IRIS K$_s$ and the WFCAM $K$ filters, and refer
the combined lightcurve bandpass simply as ''$K$''.
We have also added two photometric data points from the 2~m Fraunhofer Telescope
\citep{Hopp.2014SPIE.9145E..2DH} on
Mt. Wendelstein in Germany, taken with the 3kk optical/infrared camera \citep{LangBardl2016}.
For all photometry, the same angular size ($r = 3\arcsec$) of integrating aperture was used.
As a side note, the $K$-band image of \citet{Teixeira2012} is saturated, so that we could not use
it to extend the coverage of our light curve.

\subsection{Keck OSIRIS Spectroscopy of the Microjet}

MHO~3252~Y3 is sufficiently close to an optically visible star to allow tip-tilt
correction with the Keck~1 Laser Guide Star Adaptive Optics (LGSAO) system.
However, this tip-tilt star is at the limits of the acquisition range 
and near the practical limiting magnitude
for guide stars, so that the performance of the LGSAO
system is strongly dependent on the prevailing natural seeing.
We have obtained Keck OSIRIS observations at three epochs under varied seeing conditions.
Table~1 summarizes these observations and indicates the quality of the resulting data.
The 2012 observations were done with the coarse 100~mas scale of OSIRIS
\citep{Larkin2006SPIE.6269E..1AL}, and had the
character of ''quick first look'' observations. Nevertheless, the wavelength-integrated images in
the H$_2$ 1--0 S(1) line extracted from
these observations served as the first epoch data for the study of the microjet proper motion.
Also, the [\ion{Fe}{2}]
emission line at 1644 nm was detected in spectra taken through the OSIRIS Hn3
filter in 2012. We have not pursued 
[\ion{Fe}{2}] observations
any further since this line is faint and its emission region is spatially unresolved.
In 2015, some setup observations were taken with the 100~mas scale, which we in the end
did not use due due to problems with the PSF delivered by the AO system. The 20~mas
data from that run were of very good quality and are the basis for the detailed
study of the velocity field near the launch region of the jet.
In 2017, data were taken both in 100~mas and 20~mas scales. The seeing was noticeably
worse than in 2015, and consequently, of the 2017 data, we are only discussing the deep data in the
100~mas scale, both for jet proper motion and radial velocity studies.

The OH-Suppression Infrared Imaging Spectrograph OSIRIS
\citep{Larkin2006SPIE.6269E..1AL}
produces spectra of the pupil images formed by each of its
lenslets. The spectral resolution depends on the quality of
this pupil image formed by each lenslet, and therefore varies
across the field of view. The median spectral resolution
at the wavelength of our H$_2$ 1--0 S(1) line observations
is $\approx$~3500. The typical velocities found in protostellar
jets are of the same order as the spectral resolution of the
instrument.

The OSIRIS underwent two upgrades during this project.
The grating in OSIRIS was upgraded in 2013, leading to better efficiency
\citep{Mieda2014PASP..126..250M}
and the detector was upgraded in 2016 to a H2RG detector with better quantum efficiency.
These changes to the instrument changed the wavelength calibration and
were accounted for by updates to the data reduction pipeline.
The foreoptics, and therefore the spaxel scale and the astrometric calibration, were not affected by
the instrument upgrades.

\begin{deluxetable}{lllll}
\tabletypesize{\scriptsize}
\tablecaption{Keck OSIRIS Observations}
\tablewidth{0pt}
\tablehead{
\colhead{Date} & \colhead{100 mas} & \colhead{20 mas} & \colhead{Grating} & \colhead{Detector}\\
}
\startdata
2012 07 30 & S(1) epoch 1 & no data & old & old\\
           & [\ion{Fe}{2}] & \\
2015 05 30 & Poor PSF & Radial Velocity & new & old\\
2017 08 31 & S(1) epoch 2 & Poor Seeing & new & new\\
           & Radial Velocity & & & \\
\enddata
\end{deluxetable}

The OSIRIS data reduction pipeline (DRP) was used to extract the spectra of each individual
lenslet that defines the spatial spaxels. The extraction parameters are calibrated
with lamp spectra and stored in the form of a ''rectification matrix'' that the DRP
uses to produce a data cube with a nominal wavelength calibration.
The remaining data analysis work was done using custom IRAF scripts. The data cubes
were magnified in the spatial dimension and cleaned of cosmic ray events and other
defective pixels. 
To have an independent verification
of the wavelength solution during the observations,
we used several bright OH airglow lines that were recorded in the wavelength range of the Kn2 filter
used for our observations of the H$_2$ 1--0 S(1) line.
Two bright OH airglow lines --
9-7 R1 (2.5) at 2117.6557~nm and 9-7 R1 (1.5) at 2124.9592~nm
\citep{Rousselot2000} --
are located on either side of the H$_2$~1--0~S(1) line.
The average of those two lines is 2121.30745~nm
very close to the laboratory wavelength of the S(1) line of 2121.833~nm
\citep{Bragg.1982ApJ...263..999B}. 
For each of these two OH lines, we computed, for each spatial spaxel, the centroid of the
OH airglow line. The average of these two centroid
frames gives the wavelength calibration zero point at 2121.30745~nm, which is within the
wavelength range of the blueshifted velocity components of the observed S(1) emission.
We found that the wavelength calibration was 
dependent on the spaxel position in the field of view.
To correct measured velocities for this, we computed an image of the centroids of the S(1) emission, initially calibrated
in nominal DRP wavelength spaxels.
Subtraction of the wavelength calibration zero point, subtraction of the difference
in wavelength spaxels from that average OH wavelength to the rest wavelength of S(1),
calibration of the dispersion, conversion to velocity, and subtraction of the VLSR,
results in an image of the velocity field of the S(1) emission.

For the 
[\ion{Fe}{2}] observations, the OH airglow line produced by the transition 5-3 R1 (2.5)
at a vacuum wavelength of 1644.2155~nm \citet{Rousselot2000} served as a wavelength calibration
reference and a measure of the instrumental resolution.

\section{Results}

\subsection{$K$-band Light Curve}

The raw light curve resulting from all the $K$ band observations
is shown in Figure~2.
It is only sparsely sampled, mainly due to constraints on the observability
of the object, but also due to
some technical problems with the IRIS telescope over the course of the
5 year monitoring campaign.
Nevertheless, the lightcurve of MHO~3252~Y3 shows substantial changes in brightness with a first indication of
a possible period of $\approx$ 2.5 years between major minima.
We have used several period determination algorithms implemented in the PERANSO
software written by T. Vanmunster, each of which gave similar results for the period:
\citet{Dworetsky1983MNRAS.203..917D}: 902.32~d,
\citet{Renson1978A&A....63..125R}:  906.6~d,
\citet{Lafler1965ApJS...11..216L}: 901.5~d,
\citet{Schwarzenberg1996ApJ...460L.107S}: 904.4~d.

\begin{figure}[h]
\begin{center}
\includegraphics[angle=-90,scale=0.35]{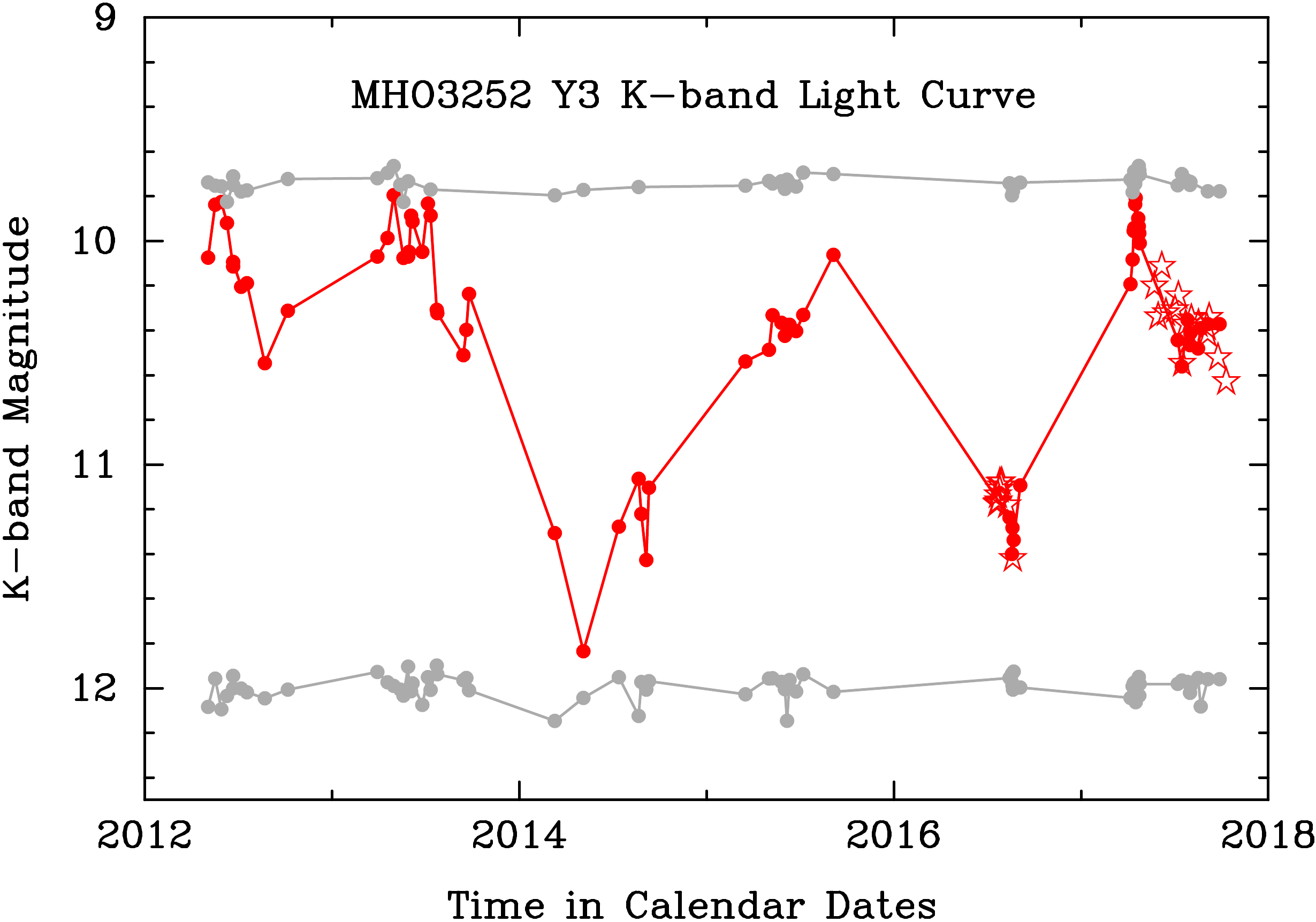}
\caption{
The light curve of MHO~3252~Y3 in the $K$~band is shown in red. The data were obtained with the 
IRIS telescope (filled circles) and UKIRT and the Fraunhofer Telescope (open star symbols) from 2012 - 2017.
For comparison, the IRIS photometry of two apparently constant stars in the field are
also shown (grey color), to demonstrate the reliability of the photometry and the typical errors
involved in the IRIS telescope photometry of similarly bright stars.
Two major dimming events, $\approx$2~mag deep, have occurred in 2014 and 2016 and
minor dimming events have been observed in 2013 and again in 2017.
}
\end{center}
\end{figure}

Somewhat naively taking the average and standard deviation of these period determinations
with different algorithms
gives P~=~904~d~$\pm$~7~d (3$\sigma$).
The light curve was phased using this period of 904~days 
and an arbitrarily chosen epoch of JD~2457857.8047 (2017 April 14)
the day during the 2017 visibility period when the brightest measurement of the first flux maximum was
obtained with IRIS.
This phased light curve shows two maxima of nearly the same brightness, and
two minima of very different depth, the first reaches down to $K\approx$ 10.5, the other down to
$K\approx$ 11.8. 
The first maximum is characterized by a sharp increase in brightness from the previous deep
minimum, while the second maximum is broader, with a slower rise from the preceding shallow
minimum. The second maximum does not occur at phase 0.5, but rather earlier, at phase $\approx$ 0.45,
so it is unlikely that the true period is 904/2~d = 452~d.

\begin{figure}[h]
\begin{center}
\includegraphics[angle=-90,scale=0.35]{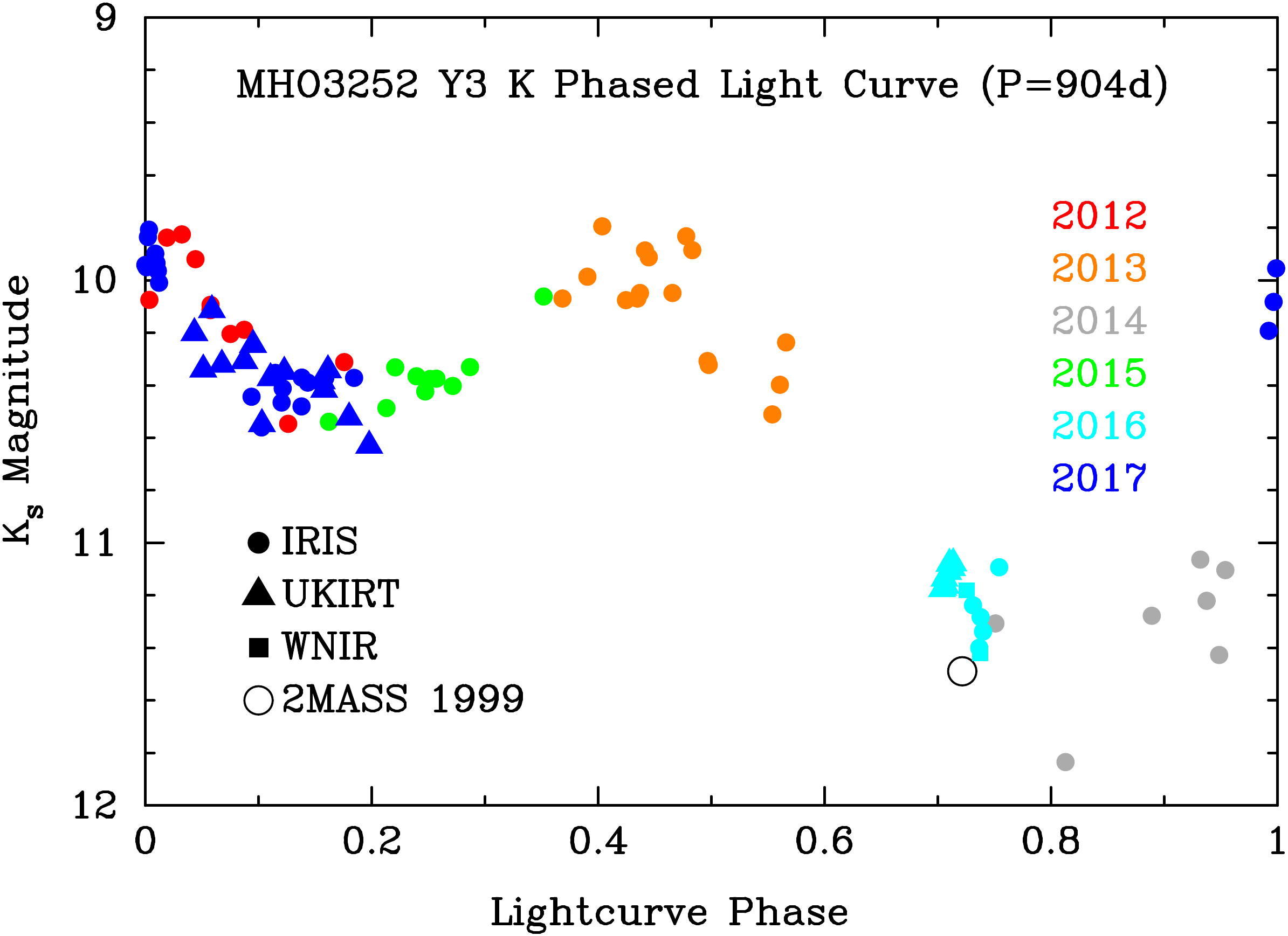}
\caption{
The light curve of MHO~3252~Y3 in the $K$~band phased with a period of 904~d.
The different colors indicate the year of the respective observations.
The phase is set zero at the epoch of the brightest measurement in 2017
on JD~2457857.8047
(2017 April 14).
}
\end{center}
\end{figure}

Figure~3 shows this phased light curve with the data points identified by year via color,
and by telescope via symbols.
The period is roughly 2.5~yr, and therefore, any part of this light curve will only
be observable every 5~yr. With the presently available data the period determination 
largely relies on matching the 2012
and 2017 data of the decline from the first maximum and on the partial overlap
of the 2015 and 2017 data on the first shallow minimum. The 2014 and 2016 data
on the deep second minimum just barely overlap. 
The fact that the 2MASS catalog \citep{Skrutskie2006} magnitude of K$_s$=11.489,
observed on 1999 April 5, i.e., 13 years before the other measurements reported here,
fits well into this phased light curve gives some confidence in the validity of
this period determination.
The 2018 visibility period will cover the broader second maximum between phase 0.4 and 0.6 again,
which was already covered in 2013. The 2019 visibility period will cover the heretofore
poorly observed broad minimum and the increase back to maximum light (phases 0.8 to 1.0).
It remains to be seen whether the light curve in the next two years will match the prediction
of this period fit to the last 5 years of data, or whether new light curve features will
emerge.

\begin{figure}[h]
\begin{center}
\includegraphics[angle=0,scale=0.45]{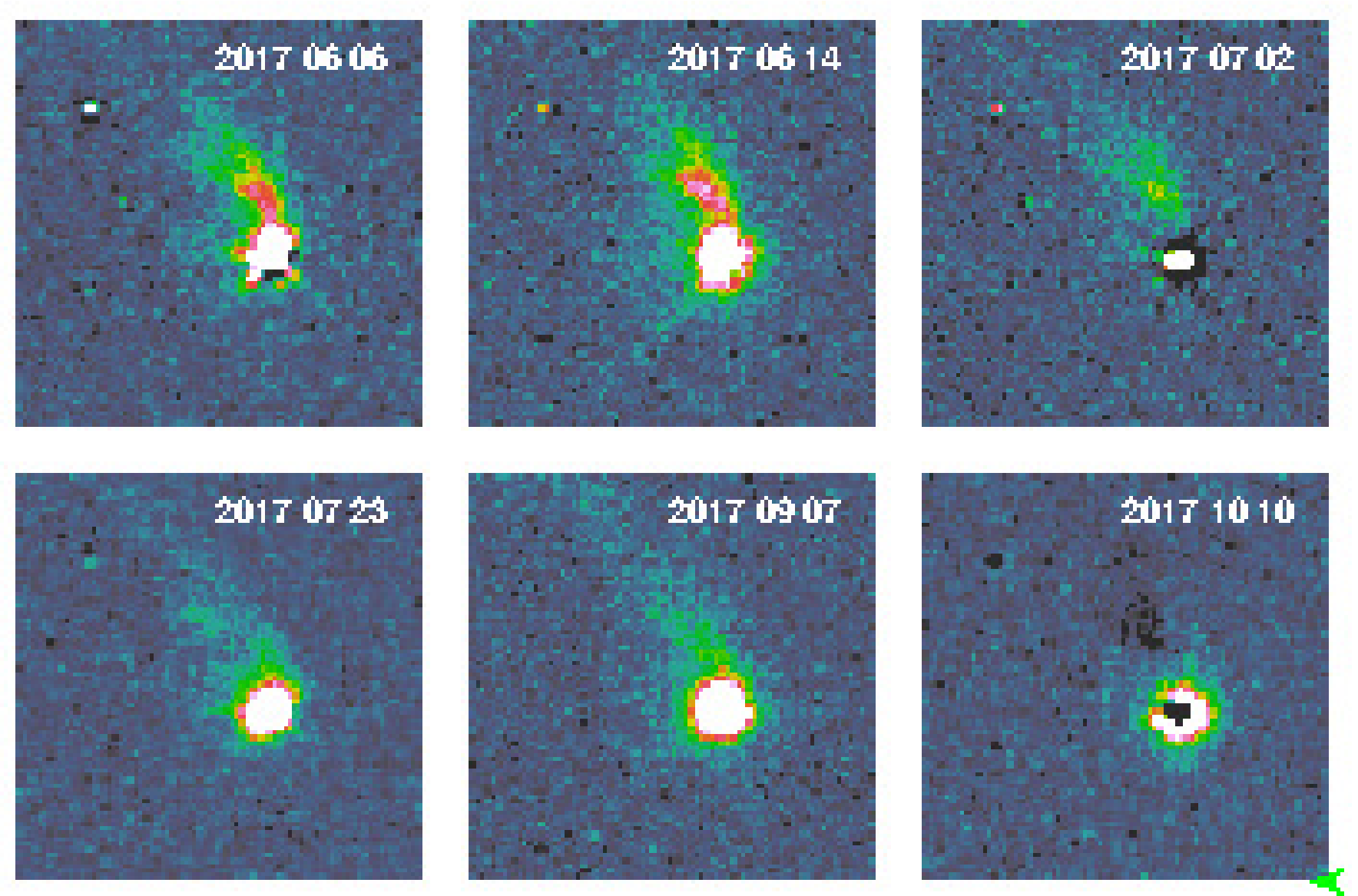}
\caption{
Difference images (30\arcsec$\times$30\arcsec) of MHO~3252~Y3 in the $K$ band 
obtained with UKIRT WFCAM during the 2017 observing season.
The average of all images in 2017 was subtracted from each individual image,
so that subtle changes in the illumination of the reflection nebula become
more noticeable. Note that the S(1) line emission of the jet is not visible in
these difference images since its flux is constant on the timescale of months
and therefore subtracts out.
}
\end{center}
\end{figure}

We have also
analyzed the broad-band $K$ images for photometric variations of the 
reflection nebula associated with MHO~3252~Y3.
In Figure~4, we show difference images at 6 selected epochs during the 2017 observing season
where the average image of all these frames obtained in that year was subtracted from
the individual images. The images are shown here in logarithmic scale in color-coded
intensity, and changes in the overall brightness of the image as well as changes
in the illumination of various components of the reflection nebula
are clearly indicated.

\subsection{Shock Front Proper Motion}

The deepest UKIRT data are summarized in Figure~1, a $H$-$S(1)$-$K$ false color image.
For Figure~1, the deep UKIRT S(1) data obtained between 2016 July 21 and 30 were combined with the sums of all the
$H$ and $K$-band UKIRT images taken in 2017 for the photometric monitoring campaign.
Molecular hydrogen 1--0~S(1) emission stands out as green features in this color composite. In particular,
it is clearly seen that MHO~3252~Y3 is the source of a system of molecular
hydrogen shock fronts that extends well beyond the highly collimated microjet close to
the driving source.
Figure~1 also shows the proper motions of S(1) shock fronts, measured from our UKIRT 2016 S(1) image and the
2011 July 16 S(1) data taken by \citet{Teixeira2012} that we downloaded from the Calar Alto Data
Archive.
After initially registering the two images with
the IRAF tasks GEOMAP and GEOTRAN,
the crosscorrelation of the two images was calculated in boxes centered on individual shock fronts using XREGISTER, and comparison
crosscorrelations were done on boxes close to the emission features but only containing stars. 
We found the pattern of cross-correlation shifts of background stars to be non-uniform across
the 2011 reference image, and therefore, we have only subtracted these cross-correlation shifts of stars locally from the
shifts measured for adjacent shock features. As a result, these proper motion vectors, 
indicated in Figure~1, are in reference to background stars, and not to the driving source
of the outflow.
We could not get any reliable proper motion measurements of the eastern bow shock. The bow shock itself was
too faint on the S(1) image from \citet{Teixeira2012} for a measurement. One set of brighter knots just
behind the bow shock has many background stars and therefore the cross-correlation technique does not
work there and we consider the proper motion vectors of these emission knots as unreliable. 

We have also measured the proper motion of the microjet emanating from MHO~3252~Y3 on the OSIRIS data cubes obtained
with the 100~mas spaxel scale of the instrument in 2012 and 2017. The raw data cubes were cleaned of spurious
pixels produced by cosmic rays and similar noise spikes. Continuum subtracted images were produced by
integrating over all wavelengths with significant H$_2$ 1--0~S(1) emission, and subtracting a continuum image
produced by similarly integrating over adjacent continuum wavelengths. The proper motions of individual
shock fronts were calculated by cross-correlation (IRAF task XREGISTER) of the 2012 and 2017 line-integrated and
continuum-subtracted data
within the integration boxes indicated in Figure~5.
The cross-correlation shift of the continuum position of the central star was subtracted from the
shifts measured for the shock fronts. The proper motion vectors in Figure~5 therefore
refer to the central star.
The proper motion of the brightest S(1) emission knot in the jet is also included
in the wide-angle image Figure~1, as a representative proper motion value for the
whole microjet.

\begin{figure}[h]
\begin{center}
\includegraphics[angle=0,scale=0.4]{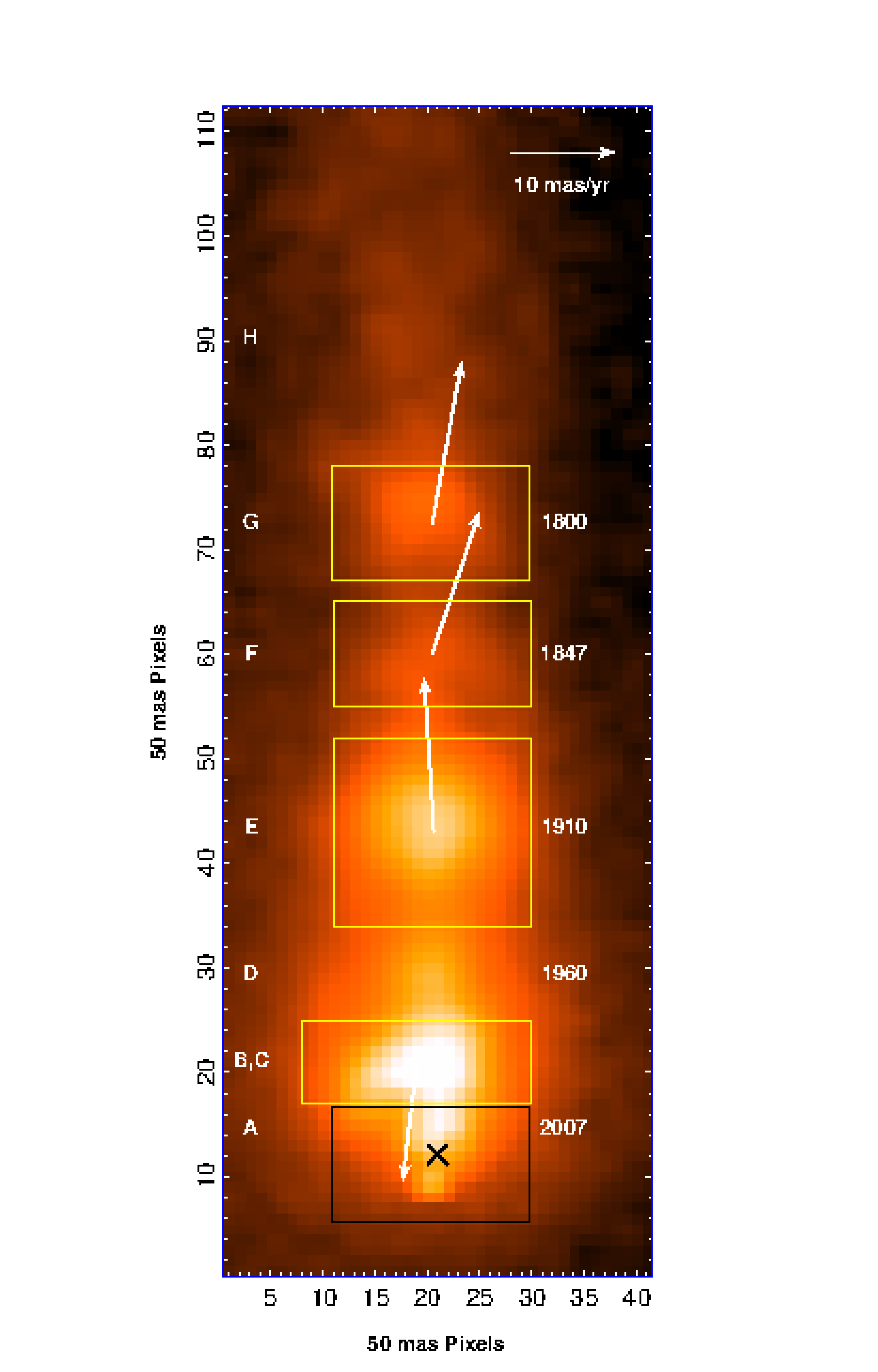}
\caption{
Keck 1 OSIRIS 100 mas/pixel data integrated over the H$_2$ 1--0 S(1)
emission in MHO~3252~Y3. 
The image was rotated by 71$\fdg$5 to show
the jet axis approximately vertical. 
MHO~3252~Y3 shows a well collimated microjet
with pronounced knot structure, indicating a variable jet ejection speed.
The proper motion vectors from 2012 to 2017, measured in the integration boxes outlined
in yellow, are indicated by the white arrows. The proper motions are relative to
the central star of MHO~3252~Y3 whose proper motion was measured in the box 
outlined in black.
The emission knots are labeled in the same way as in Figures 6 and 7 and
the approximate ejection date of each emission knot calculated from its distance
and the average proper motion is indicated.
}
\end{center}
\end{figure}

The epoch difference between the two OSIRIS 100~mas data sets from 2012~July~13  (JD 2456121.5)
and 2017~August~31 (JD 2457996.5) is 1875~d = 5.133~yr
For the best defined S(1) emission knot, 
labeled ''E'' in Figures~5 and 6 
we measured
a proper motion of 14.784~$\rm mas~yr^{-1}$
or 30.56~\rm km~s$^{-1}$ tangential motion at the adopted distance
of 436~pc.

The median radial velocity (see Section 3.3) of this knot was 
measured over the same integration box as was used for the cross-correlation
for the proper motion measurements and was calculated relative to the
motion of the molecular cloud in the -- unproven -- assumption that this
is a good proxy for the motion of the central star.
This measured radial velocity of the knot of -86~kms$^{-1}$ relative to the star,
implies
that the jet is oriented only 19$\fdg$6 inclined against the line of sight.

\subsection{OSIRIS Radial Velocity Measurements}

We are presenting the radial velocity information from the "wide field" 100~mas spatial scale data in Figure~6,
and in Figure~7 for the "high resolution" 20~mas~scale of OSIRIS. 
In both figures we have coded the radial velocity by color, and have multiplied
the resulting RGB color image by the line-integrated intensity, such that a zero
intensity shows as a white color and the color saturation level indicates intensity. 
The reason for this method of presenting the velocity data is to prevent 
regions of low flux from dominating the velocity diagram with high-noise pixels. 

Our velocity data are presented relative to the local standard of rest
velocity (VLSR) and all measured velocities are blueshifted relative to this velocity. 
From the $^{13}$CO measurements of \citet{Nakamura2017}, the velocity of the Serpens~S
molecular cloud and of the nearby W~40 complex
is 7~km s$^{-1}$ relative to VLSR and they argue that both the angular proximity and essentially identical
velocity strongly suggest a physical connection of the two clouds.
If we assume that the driving source of MHO~3252~Y3 has the same velocity as the
surrounding molecular cloud, the jet velocities are,
7~km s$^{-1}$ more blueshifted relative to the star.

We distinguish two velocity features:\ 
1. The well collimated, highly blueshifted velocity component
that we will simply refer to as the ''jet''. The jet has an ''onion-like'' nested velocity
structure with a fast moving core and relatively slower envelope.\
2. A spatially much wider, distinctly lower velocity component indicated in red color
if Figure~7, and visible mostly in the lower velocity panels of Figure~8, that we
refer to as the broad low velocity component (BLVC) in the following.

\begin{figure}[h]
\begin{center}
\includegraphics[angle=0,scale=0.8]{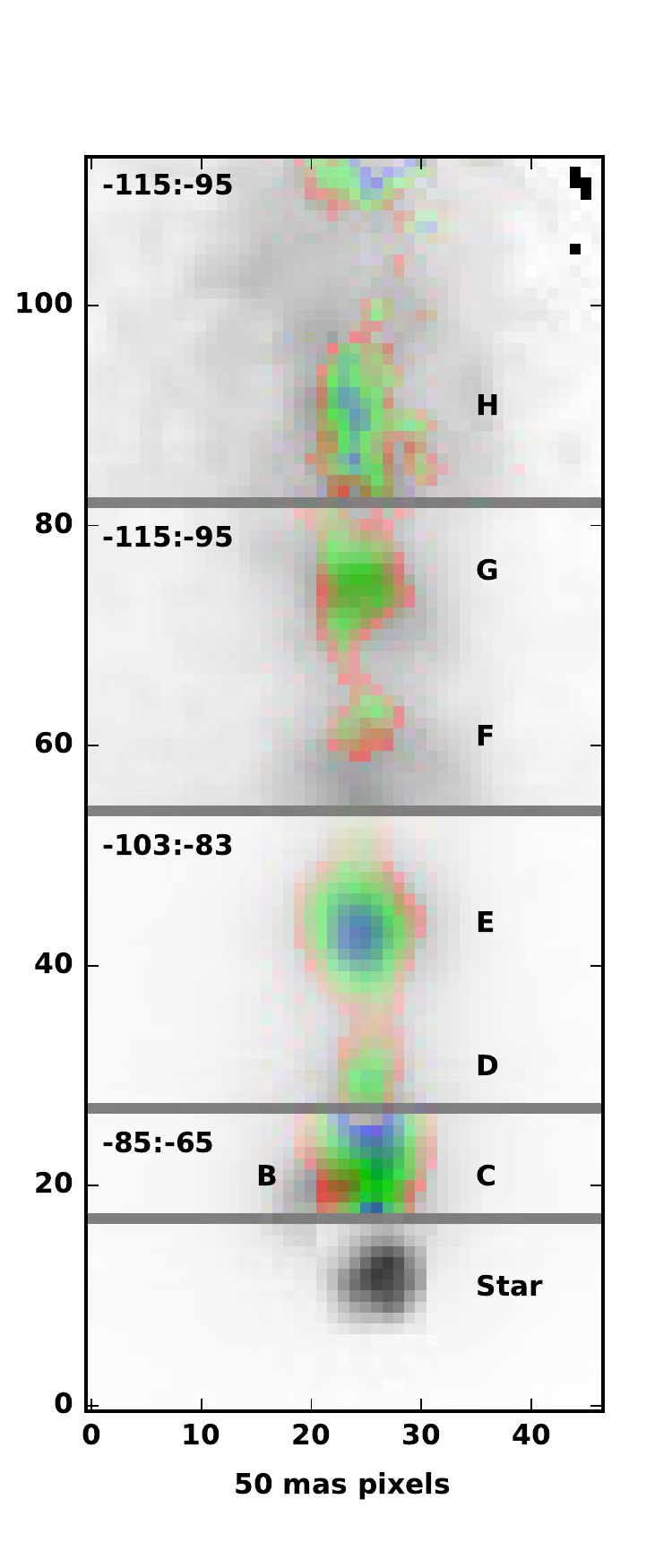}
\caption{
Velocity (coded by color) times flux (coded in color density) of MHO~3252~Y3 in H$_2$ 1--0 S(1) based on the 
OSIRIS low-resolution data
that are displayed here on 50~mas pixels. 
The image was rotated by 71$\fdg$5 to show
the jet axis approximately vertical. 
The velocity range displayed in the full RGB range
is always 20~km$^{-1}$ VLSR, but the velocity window has been shifted in the three different regions
of the figure, separated by grey lines, so that the display parameters could be individually adjusted to
the different intensities and velocity ranges of the emission features, so that the velocity structure of the jet is shown optimally. 
A continuum image of the star has been added in grey-scale as a position reference.
The highest velocities in the jet are near its center axis, while its envelope shows lower velocity
due to the entrainment of the wider low velocity disk wind, the BLVC.
}
\end{center}
\end{figure}

\begin{figure}[h]
\begin{center}
\includegraphics[angle=0,scale=0.6]{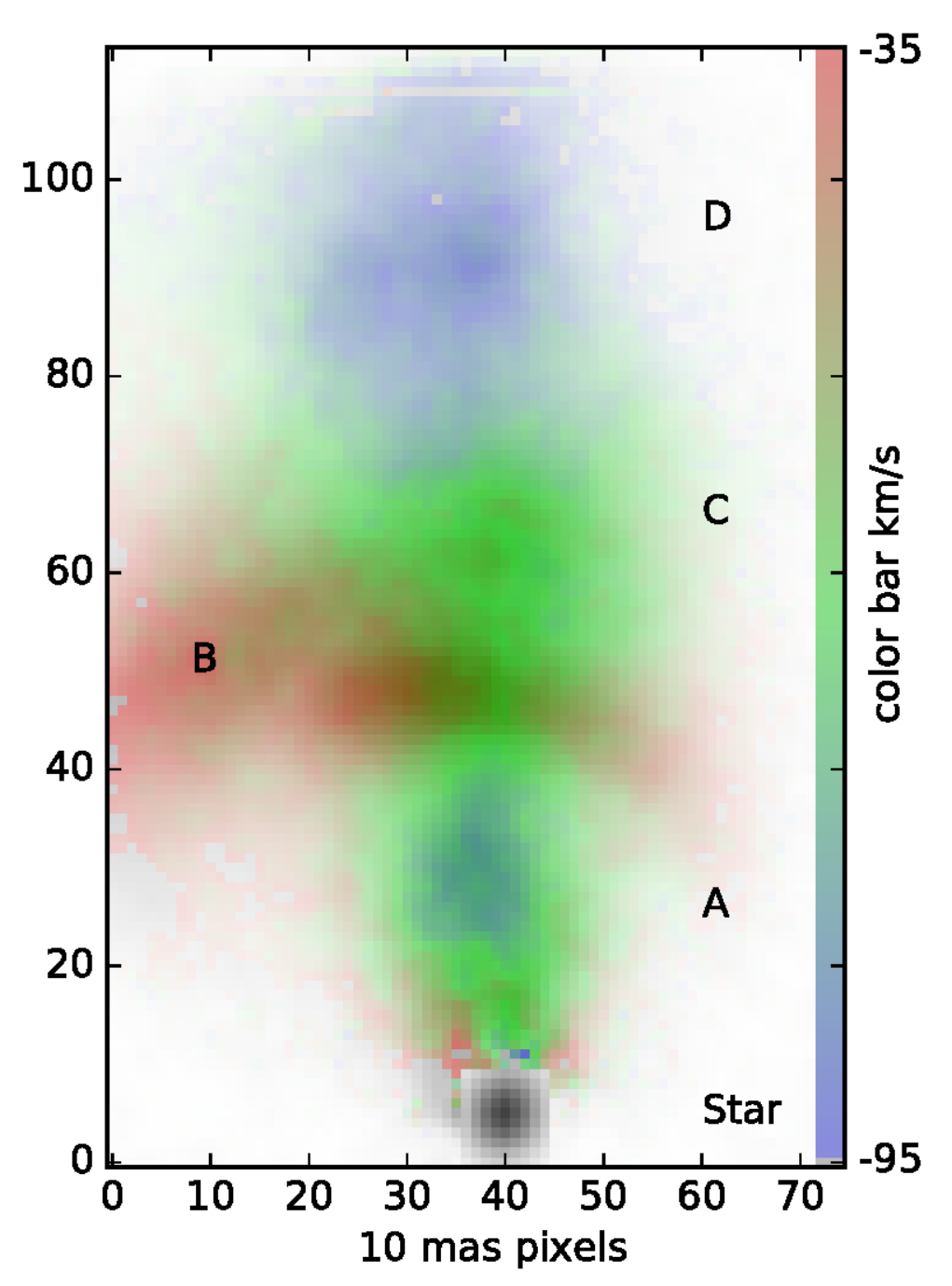}
\caption{
Velocity (coded by color) times flux (coded in color density) of MHO~3252 Y3 in H$_2$ 1--0 S(1) 
based on the high-resolution data (OSIRIS 20 mas scale)
that are displayed here on 10 mas pixels. 
The image was rotated by 71$\fdg$5 counter-clockwise to show
the jet axis approximately vertical. 
The velocity range is 35 $\rm km^{-1}$ - 95 $\rm km^{-1}$
as indicated by the color bar. 
A continuum image of the star has been added in grey-scale as a position reference.
The velocity range of this figure is 3 times that of Figure 6, reflecting the 
large difference of the velocities of the jet (HVC) and the BLVC.
}
\end{center}
\end{figure}

\begin{figure}[h]
\begin{center}
\includegraphics[angle=0,scale=0.40]{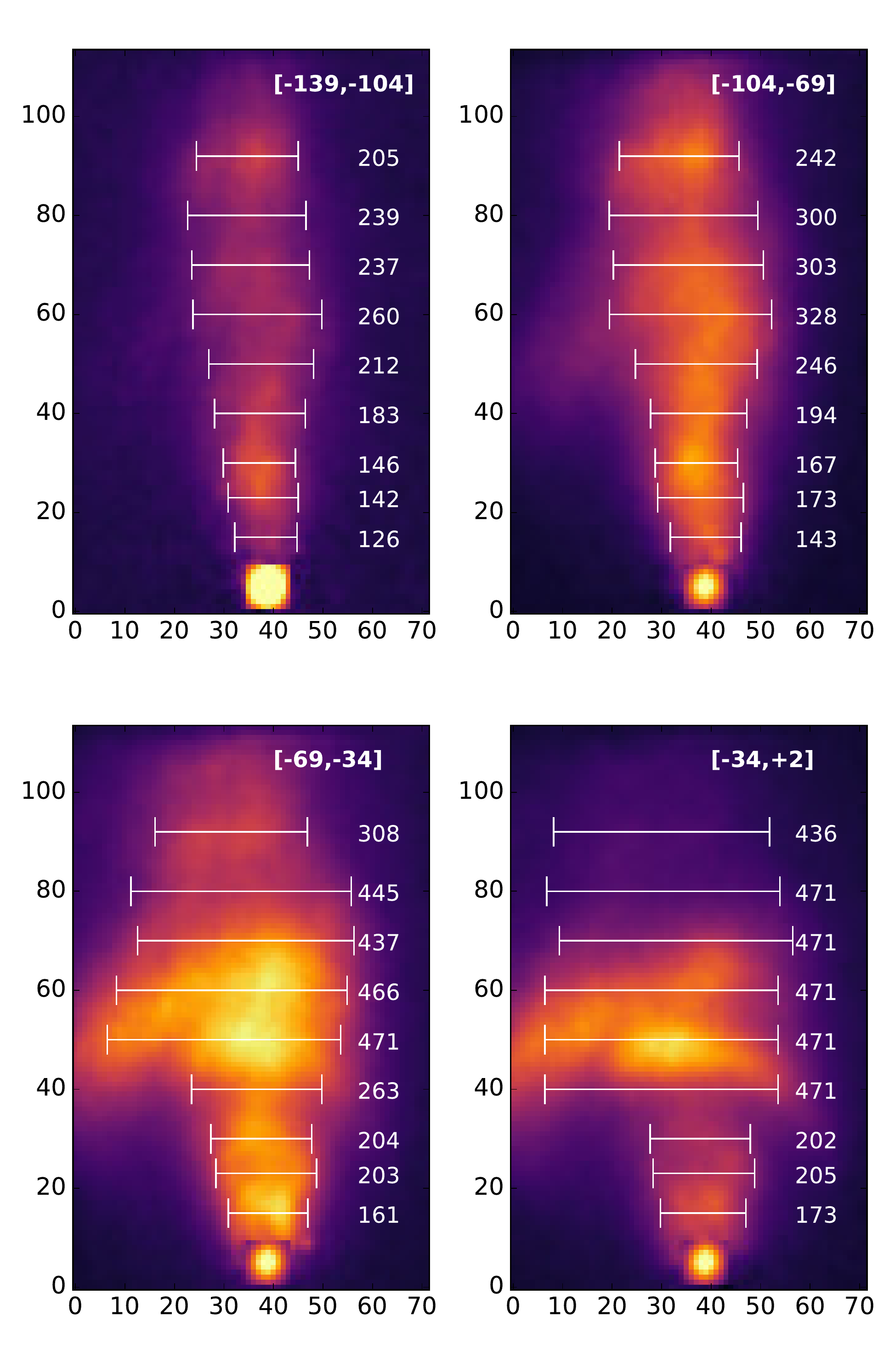}
\caption{
Four wavelength channels of the 2015 OSIRIS data in the 20 mas scale, displayed
here on 10 mas pixels. 
These channel images were continuum subtracted, but 10\% of the flux of the continuum image
has been re-inserted at the position of the driving star to identify its position.
Even ignoring the BLVC, several areas in the jet show asymmetric features with
wavelength that create a rotational component to the velocity field. However, the velocity field is far more
complicated than simply a rigid rotation of the microjet.
}
\end{center}
\end{figure}

Individual wavelength planes of the continuum-subtracted data cube of the 2015 data
in the 20 mas spaxel scale are shown in Figure~8. The spatial scale has been magnified
by a factor of two, so the axes tickmarks indicate 10 mas pixels. The velocity range
covered by each wavelength plane is given relative to the LSR. The motion relative
to the molecular cloud, and presumably relative to the star, is 7 km~s$^{-1}$ more
blueshifted. The width of the jet was measured in stripes 50 mas in size in the direction
of the jet axis (vertical in Figure~8) by fitting a Gaussian profile to these stripes.
The symbols in Figure.~8 give the FWHM, i.e., 2.355$\times\sigma$, for compatibility
with most other studies of jet width and collimation.
The fastest approaching (blueshifted) components of the jet show the narrowest width,
while slower approaching components are wider. This is consistent with the ''onion-layer''
model of a jet where the outer layers of the jet get slowed down by turbulent 
interaction with a slower-moving wind component, as was discussed in the context
of the DG Tau jet by 
\citet{Agra-Amboage.2011A&A...532A..59A} and
\citet{2016MNRAS.455.2042W}.

We have indicated the position of the central star by inserting the continuum image
of the star at the correct position and with 10\% of the original intensity.
At the closest distance from the star that we could measure the FWHM, the jet already
appears collimated and expands from that position monotonically with a small opening
angle, until the jet passes through the broad low-velocity component. Beyond this
BLVC, the FWHM measurements are more varied, possibly indicating an interaction
with that low-velocity component.

\newpage
\subsection{Detection of [\ion{Fe}{2}] Emission}

During the initial observations in 2012, we also searched for 
the 1644.002~nm \citep{Aldenius2007} emission of
[\ion{Fe}{2}], but did not
detect any emission from the microjet in that high-excitation line.
However, faint, blueshifted [\ion{Fe}{2}] emission was indeed
detected at the position of the driving source.
The spectrum in Figure~9 was extracted from a continuum-subtracted
data cube in a 300~mas
diameter aperture centered on the continuum source.
This indicates that close to the launch region of the jet, unresolvable
by our observations, Fe has been released from dust grains into the gas
phase in sufficient numbers, and the higher excitation conditions exist to form the
[\ion{Fe}{2}] line emission in an otherwise very low excitation jet.
Due to the low signal-to-noise ratio of this detection, we did not do any
further analysis of this finding.
Using the nearby 5-3 R1 (2.5) OH airglow line as a wavelength calibrator,
we present the extracted spectrum in terms of velocity shift against
the [\ion{Fe}{2}] VLSR wavelength in Figure~9. The [\ion{Fe}{2}] emission is blueshifted
relative to VLSR and the Serpens S systemic velocity of 7 km s$^{-1}$ 
by about the same amount as the S(1) velocity components in Figure~7.
The line width of the [\ion{Fe}{2}] emission is not significantly different
than that of the spectrally unresolved OH airglow line, indicating that
this [\ion{Fe}{2}] emission originates in only one velocity component.

\begin{figure}[h]
\begin{center}
\includegraphics[angle=0,scale=0.45]{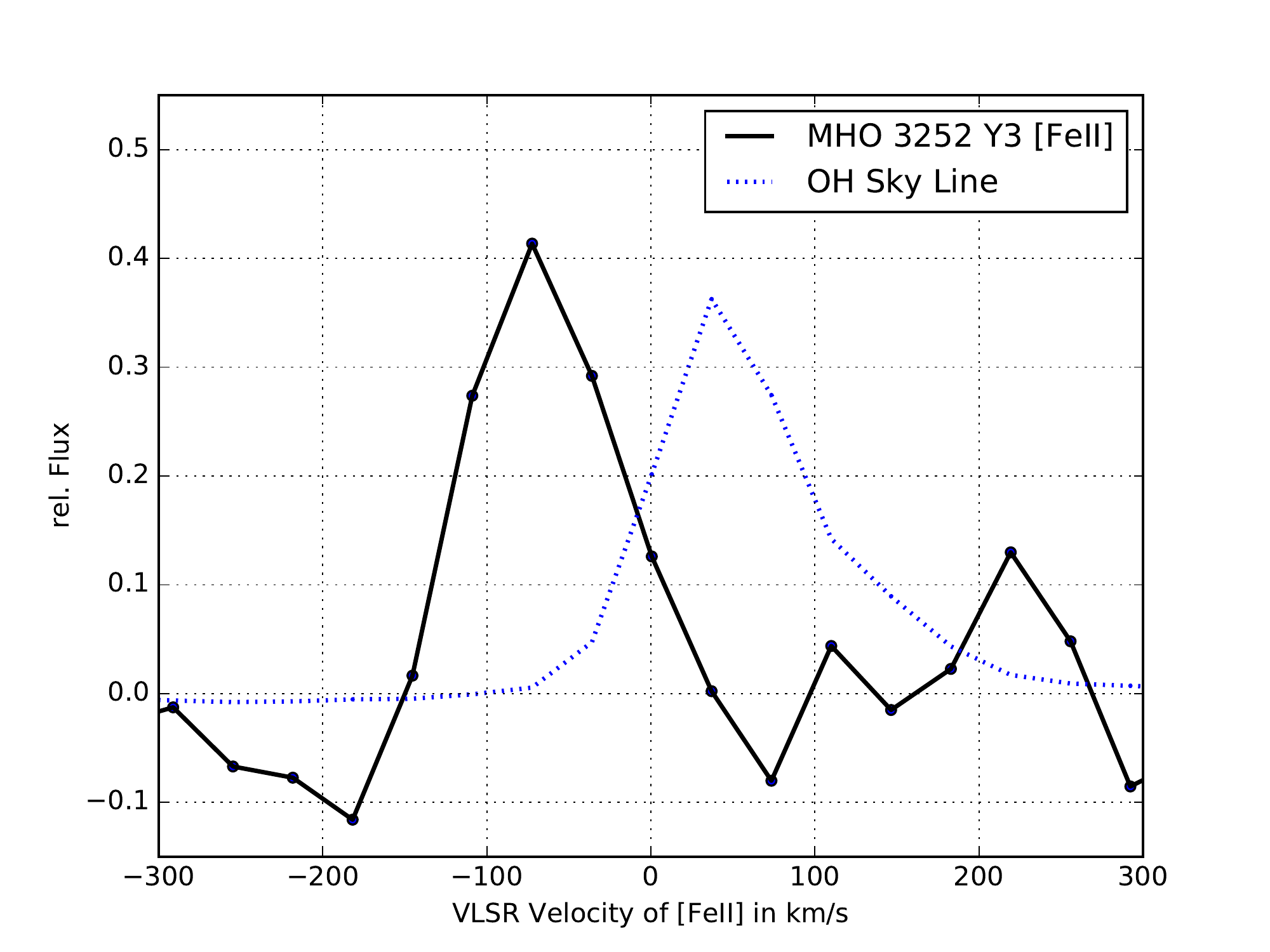}
\caption{
Spectrum of [\ion{Fe}{2}] emission in a 300 mas diameter aperture 
at the position of the MHO~3252~Y3 central star.
For comparison, the scaled profile of the nearby OH 5-3 R1 (2.5) OH airglow
line is shown, demonstrating that the [\ion{Fe}{2}] line is not significantly
broadened.
}
\end{center}
\end{figure}

\section {Discussion}

\subsection{Photometric Variability}

Young stellar objects are almost all measurably variable and the diversity
of variability types has recently been reviewed by
\citet{Hillenbrand.2015ApJ...808...68H}.
The most common type of variability of young stars is the rotational modulation
due to star spots on the surface, leading to short-period small-amplitude variations.
MHO~3252~Y3 is clearly not of this nearly ubiquitous type of variability.
Based on Spitzer Space Telescope and CoRoT (Convection, Rotation, and planetary Transit) satellite, 
as well as ground-based monitoring data \citet{Stauffer..2015AJ....149..130S},
\citet{Stauffer..2014AJ....147...83S}, and
\citet{Cody-2014AJ....147...82C}
have studied young stars in their accretion phase in the NGC~2264 star-forming region.
Among the stars with periodic light curves, they distinguish
''bursters'' and ''dippers'', the former erupting from a quiescent
brightness level, the latter dipping down from a normally bright state.
A direct comparison with our observations is not straightforward, since
their studies concentrated on objects with much shorter periods -- a few days -- than our object.
More importantly, MHO~3252~Y3 does not have a well-defined ''normal'' brightness level, neither
at maximum nor at minimum brightness.
The light curve of MHO~3252~Y3 shows that its brightness is between $\approx$ $K$=9.8 and 10.6
for most of the time, and shows one much deeper minimum down to $\approx$ $K$=11.8.
Also, we note that the two maxima of the phased light curve (Figure~3) reach the same magnitude: $K$=9.8,
while there is a first shallow minimum down to $K$=10.6 (phase 0.18), and a second,
deeper minimum down to $\approx$ $K$=11.8 at phase 0.8. The more
common state of this object is near maximum brightness and this maximum is better defined.
We therefore interpret this object as a ''dipper'', and explain
the variability of the central star by variations of the extinction by a thick disk structure
surrounding the central star and partially obscuring 
the line of sight. The continuum flux from MHO~3252~Y3 probably has a substantial and possibly dominant
contribution from scattered light when integrated over the aperture (3$\farcs$0) of our
photometric monitoring observations. We have to expect that these variations in the extinction are
individual to different scattering paths, i.e., for different parts of the reflection
nebula. We have indeed observationally confirmed (Figure~4) that the changes in the
reflection nebular are not uniform, but that different parts of the reflection nebula
vary independently of each other. 

Variability of the light distribution in young reflection nebulae has been observed for
a long time, e.g., \citet{Hubble..1916ApJ....44..190H} reported on
the variability of the cometary reflection nebula NGC~2261, illuminated by R~Mon,
and discussed other earlier observations of similar objects.
More recently, observations of infrared reflection nebulae,
e.g., \citet{2009AJ....137.3501H}
and \citet{Connelley..2009..AJ..137.3494C} have shown that variability of the light distribution in such
objects is rather common.

In order to increase the signal-to-noise ratio for a study of the color variations associated with the photometric variability,
we have averaged all UKIRT WFCAM observations from the summer of 2016 near minimum brightness, and the first six
observations from 2017, just after the maximum, in the three filters $J$,$H$, and $K$.
Figure~10 shows that the amplitude of the variation measured simultaneously in $J$, $H$, and $K$
is wavelength dependent such that shorter wavelengths vary more, but
that quantitatively, this wavelength dependence is  flatter than that expected from varying levels of interstellar extinction 
in front of an object of constant color.
We have plotted the expected values of the variation amplitude in $J$ and $H$,
when normalized to $K$, in Figure~10 for two extinction ''laws''. The extinction law of \citet{Rieke1985ApJ...288..618R}
and a newer determination of the extinction law in the direction of the Galactic center
\citep{Nishiyama2009ApJ...696.1407N}, both referring to ice-free interstellar lines
of sight, predict more variation with wavelength than we observe.
This is quite in line with
the expectation of a flatter extinction law for the larger grains typical of protoplanetary disks.

\begin{figure}[h]
\begin{center}
\includegraphics[angle=-90,scale=0.35]{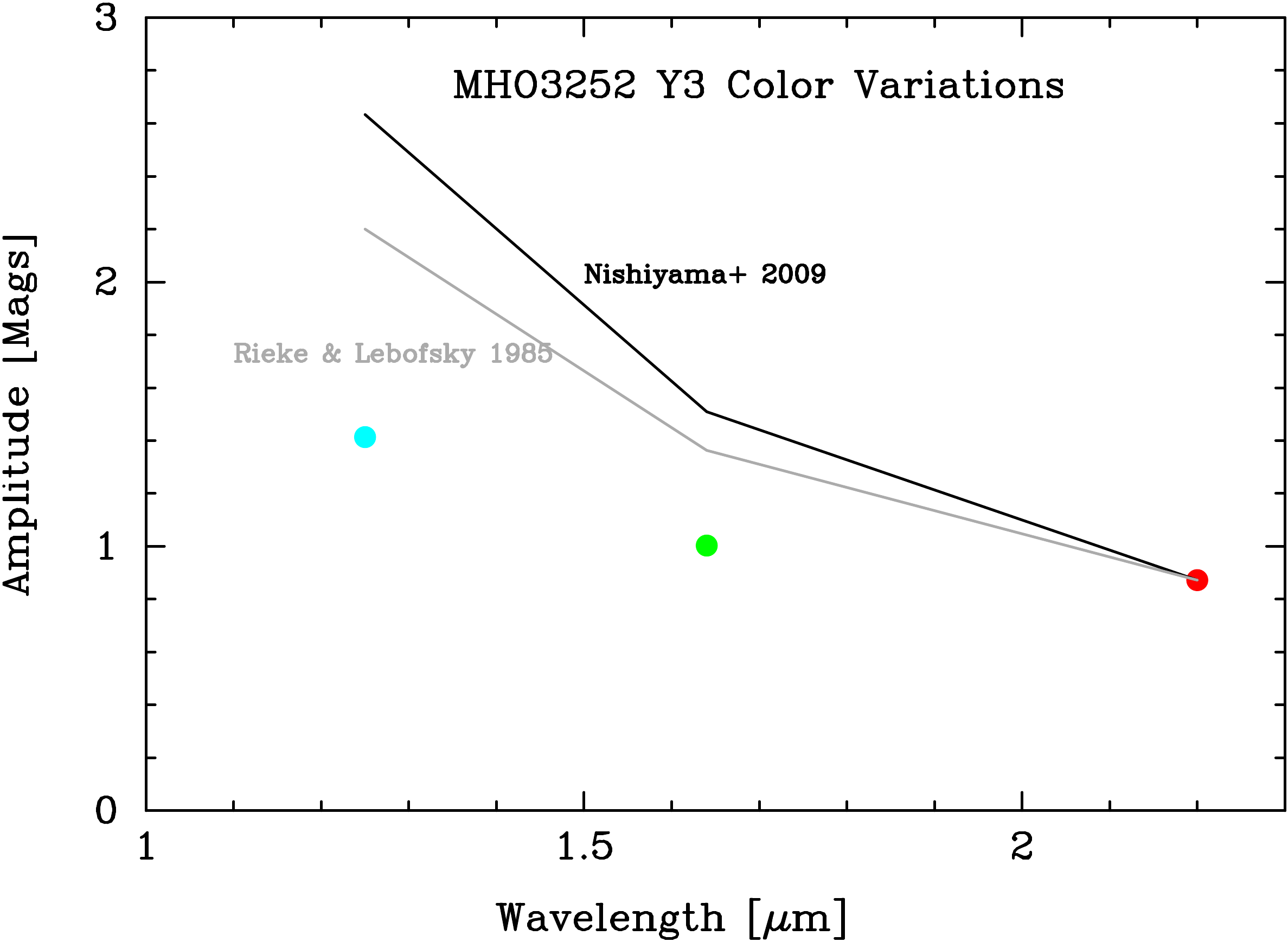}
\caption{
The amplitude of simultaneous integrated flux variations of MHO~3252~Y3 in the $J$, $H$, and $K$ filters 
is shown as blue, green, and red filled circles, respectively. These color variations cannot be explained by the two interstellar extinction laws
shown as solid lines.
Rather, the colors indicate that the absorbing dust in MHO~3252~Y3 has a flatter extinction law 
characteristic of a larger particle size distribution.
}
\end{center}
\end{figure}

\subsection{Morphology and Proper Motion of the Jet}
The measurement of shock front proper motion in Figure~1 allows to identify
the driving sources of the two outflows in this field.
West of MHO~3252~Y3, the S(1) emission feature labeled MHO3252r by \citet{Teixeira2012}
has a strong proper motion at P.A. $\approx${260}$\arcdeg$, clearly identifying this as the
western, redshifted lobe of the outflow from Y3.
The eastern lobe is morphologically connected to the outflow cavity scattered light
seen in continuum $K$ band by faint S(1) line emission, and the microjet is centered in this outflow cavity. 
This eastern lobe ends
in a typical bow shock (near the left edge of Figure~1), but we could not measure any proper motions of this bow shock
because the 2011 image of \citet{Teixeira2012} did not record this emission with
sufficient signal to noise ratio to allow cross-correlation with our 2016 UKIRT image.
The brighter emission
feature behind the western bow shock labeled MHO~3252b by \citet{Teixeira2012} 
is superposed on several faint
background stars, interfering with the cross-correlation method for the proper motion
measurement. We show the resulting proper motion vectors in Figure~1, but do not
believe that these measurements are valid. Despite this inconclusive proper motion
measurement
this feature is morphologically clearly part of the Y3 outflow
and coincides
with blueshifted CO emission mapped by \citet{Nakamura2011}
as had already been pointed out by \citet{Teixeira2012}. 

We have also measured the features MHO~3251b and MHO~3255, north and south of the Y3 outflow, respectively, that coincide with blueshifted
CO emission in the map of \citet{Nakamura2011}  and find that their
proper motions strongly suggest an origin in the Serpens~S Main cluster to 
the north of MHO~3252~Y3,
probably in the deeply embedded object P3 that \citet{Teixeira2012} had
already identified as the likely driving source of the large outflow originating in
the Serpens South Main cluster and moving towards the south.

In the proper motion data on the microjet in Figure~5, based on the OSIRIS 100~mas scale
measurements of 2012 and 2017, the
measured shift of the photocenter in the emission region ''B''
is towards the star, a result that cannot be interpreted
as the proper motion of a physical element of the outflow. This particular
region close to the driving source shows two distinct radial velocity components, the highly blue-shifted
microjet -- the HVC -- and a spatially wider, less blue-shifted BLVC. The wavelength-integrated
image, on which the proper motion measurement is based, does not distinguish these
two components, and the photocenter shift most likely is caused by changes in
the relative intensity of the microjet HVC and the BLVC, rather than physical
motion of either of these components. We do not have a high spatial resolution
data set with the 20 mas scale of OSIRIS at the early (2012) epoch. A further investigation of motion and
photometric changes of those two radial velocity components will require
additional high spatial resolution data in future years.

For each well defined emission knot, we have calculated the age by dividing
the distance from the star position (indicated by the X symbol) by the
average proper motion of 15~$\rm mas~yr^{-1}$. We have listed the resulting approximate
dates of the shock front ejection in Figure~5. The shock front closest to the
star at a distance of only 150~mas, was therefore ejected approximately in 2007,
i.e., before our monitoring campaign started. Therefore, if the ejection
of this latest knot was associated with a photometric outburst, it happened
just before the discovery of this object and was not recorded.
Photometric outburst associated
with modulations of the outflow speed should repeat themselves
about every 50 - 60 years, much longer than the photometric periodicity found here.

The microjet S(1) emission observed
in the UKIRT image extends for about 9\arcsec, much beyond the coverage of our
OSIRIS 100 mas observations shown in Figure~6. If we assume that the microjet
as a whole has the typical proper motion of those knots that we have measured,
15 mas yr$^{-1}$, its kinematic age is 600 yr.

\subsection{Nested Velocity Structure in the Microjet}

The MHO~3252~Y3 microjet is oriented only 19$\fdg$6 inclined against the line of sight
and the radial velocity measurements are therefore very sensitive to the 
differential velocities within the jet. 
The approaching microjet has the highest blueshifted velocities in Figure~6,
and therefore is shown in blue color. Intermediate velocities are shown in green. 
Of all the 
blueshifted velocity components, the relatively lowest velocity is shown in reddish color. 
The highest
velocity blueshifted emission has the narrowest spatial width, as indicated
by the blue color in Figures 6 and 7, and by the FWHM of different velocity
channels in Figure~8. 
We identify
this highest velocity component with the center of the microjet where the material
had the least turbulent interaction with surrounding slower wind components,
and therefore maintains the highest velocity. 
The highest blueshifted velocity regions in the microjet are, however, surrounded by faint
emission of lower velocity, shown as a faint reddish halo around it. Alternatively, 
Figure 8 shows that the lower velocity channels show a wider FWHM.

We interpret this nested velocity structure surrounding the microjet
as an envelope of lower velocity material, most likely
the result of interaction of the microjet with a lower velocity disk wind.
This velocity structure is also visible at the base of the jet in the
high spatial resolution data of Figure~7, where the center of knot A
shows blue color, surrounded by an envelope of slower material, shown
in green, with a faint halo of low velocity material indicated in red.

Finally, the broad, relatively low velocity component (BLVC) that
stands out in the proper motion study as having an apparent reversed motion
is also morphologically and kinematically distinct 
in Figures 7 and 8.
We will discuss this feature in subsection 4.5.

As already outlined in the Introduction, we are comparing our results
on MHO~3252~Y3 to those on DG~Tau.
DG~Tau is a T~Tauri
star in the transition phase from Class~I to Class~II \citep{Pyo.2003ApJ...590..340P}
\citep{White.2004ApJ...616..998W},
without a detected CO outflow and a faster,
higher excitation microjet primarily detected in [\ion{Fe}{2}] 
\citep{White.2014MNRAS.442...28W} and with H$_2$ emission only detected 
directly near the collimating disk cavity. DG~Tau is therefore probably
less embedded and
more evolved than MHO~3252~Y3.

Our findings for the MHO~3252~Y3 microjet are very similar to those for the well studied DG~Tau
microjet, where evidence for similar lateral entrainment was discussed by 
\citet{Pyo.2003ApJ...590..340P} and
\citet{White.2014MNRAS.442...28W}.
The main difference is that the DG~Tau microjet is primarily detected in the
higher excitation [\ion{Fe}{2}] line at 1644~nm while the
MHO~3252~Y3 microjet is only detected in the lower excitation H$_2$ 1--0 S(1)
line. 

The formation of an intermediate velocity, shock excited layer at
the interface between the high velocity microjet and the low velocity disk
wind is otherwise identical to the
velocity structure discussed and theoretically modeled by 
\citet{2016MNRAS.455.2042W} for DG~Tau (specially their Figure~1).
The salient point in their model is that protostellar microjets
have an ''onion-like'' velocity structure where the highly collimated
fast jet turbulently entrains material from the low velocity disk wind
component, forming a turbulent mixing layer of intermediate velocity
that is shock-excited. It has been argued by 
\citet{2016MNRAS.455.2042W}
that this mechanism, usually referred to as lateral
entrainment in distinction to the frontal entrainment happening in
leading bow shocks, does indeed operate even for very fast jets
such as that in DG Tau when the effect of magnetic fields is 
accounted for.
The same nested velocity
structure has been found in almost all jets 
studied in sufficient detail, as reviewed by \citet{Frank..2014..prpl.conf..451F}. 

\subsection{Jet Collimation}

In Figure~8
we show four individual velocity channel images  of the high spatial
resolution (OSIRIS 20 mas scale) data and indicate the 
measured FWHM of the microjet emission measured in 5 pixel high stripes. 
Due to its orientation towards the observer, the MHO~3252~Y3 microjet
is not well suited for a study of its collimation.
Therefore, our high spatial resolution velocity map in Figure~7 and the individual
velocity channels in Figure~8 only show that the microjet is essentially collimated
at the closest position to the star that we could get a FWHM measurement for, 70 mas
from the central star, and
from there only expands with a narrow expansion angle. 

At the distance of MHO~3252~Y3, 70 mas distance before full collimation corresponds to a projected distance
of 30.5 au from the star. 
The perspective shortening of the jet due to its angle of only 19\fdg6 against
the line of sight leads to a shortening by a factor of $\approx$~2.8.
Applying this de-projection, the jet appears collimated 85 au along its
axis from the star.
At that position, the measured FWHM in the -104 to -69~kms$^{-1}$ velocity
range is 143 mas or 62 au wide. 

Model for the collimation of jet have been presented by
\citet{Garcia.2001A&A...377..609G} 
and 
\citet{Cabrit.2007A&A...468L..29C} have presented radio interferometric
results on the HH 212 jet and compared those to older results on
the collimation of Class II jets by
\citet{Claussen.1998ApJ...507L..79C}, 
\citet{Dougados.2000A&A...357L..61D}, and
\citet{Woitas.2002ApJ...580..336W}.
Our results are broadly consistent with those earlier results.
The MHO~3252~Y3 microjet must be essentially collimated about 85~au from 
the central star along
the axis, at which point it is already 62 au wide.
Due to larger distance of MHO~3252~Y3 and orientation towards the observer, our
data cannot resolve the collimation region itself.

\subsection{The Broad Low Velocity Component (BLVC)}

While essentially all the S(1) line emission of the MHO~3252~Y3 microjet is blueshifted
we have two components of distinct velocity, spatial extent, morphology, and proper motion.
The BLVC has less than half of the radial velocity
of the fastest components of the well collimated microjet. 
If we assume for the moment that the BLVC was generated
in an eruptive event, any element of the better collimated microjet component with about double the 
velocity that was ejected at the same time would have had time to travel just about to the top
of the field of view of the 2016 OSIRIS 20 mas observations shown in Figures 7 and 8. However, no unusual 
features are visible in the jet at that location. We conclude that the BLVC can only be
the product of a disk instability event if this event does not at the same time
affect the generation of the HVC microjet.

We again compare our results to the
similar, but better studied case of the YSO jet in DG~Tau.
A stationary shock at the basis of DG~Tau jet has been observed in [\ion{Fe}{2}] by \citet{White.2014MNRAS.442...28W}
and interpreted as a ''recollimation'' shock from interaction of the
outflowing material with the outflow cavity that collimates the outflow into
a narrow jet. Observations of the emission of H$_2$ S(1) in DG~Tau
by \citet{Agra-Amboage.2014A&A...564A..11A}
show emission over a circular
area of $\approx$500 mas diameter at the basis of that jet.
They presented four different models to explain a spatially broad, low velocity
S(1) emission region at the base of the DG Tau jet.
Considering the closer distance of DG~Tau, this is about 1/3 the linear
size of the low-velocity S(1) emission region in MHO~3252~Y3.
Moreover, the MHO~3252~Y3 BLVC has the shape of two
overlapping arcs, not a circular shape as in DG~Tau.
The BLVC in the MHO~3252~Y3 outflow 
is seen $\approx$ 450 mas from the star, a projected distance of $\approx$200 au, and
a de-projected distance of over 560 au, which is much farther in linear distance than
the low velocity features in the DG Tau microjet. 

A possible, though speculative, explanation for this wide low velocity
feature is a particularly strong internal shock within the jet that led
to sideways splattering of jet material that subsequently interacts with
the surrounding slower disk wind or even ambient molecular material.
In their detailed study of several optically visible jets, including
HH1-2 and HH 46-47, \citet{Hartigan.2011ApJ...736...29H} 
have found numerous cases of knots changing brightness, and also cases
of knots and shock fronts expanding over the course of 13 years.
In the outflow emanating from the binary T~Tauri star system XZ~Tau,
\citet{Krist2008}
have found a string of knots forming a jet, surrounded by a series of bubbles that interact
with each other to form a broader envelope around this central jet.
Similarly, in the outflow of the Class I embedded object NGC~1333~SVS13,
\citet{Hodapp2014ApJ...794..169H} found a high-excitation jet, traced in
[\ion{Fe}{2}] emission surrounded by wider, lower excitation shock fronts
traced in the H$_2$ 1--0 S(1) line, the youngest of which appear to be
near spherical expanding bubbles. 
The jet and bubbles in NGC~1333~SVS13 were numerically modeled by
\citet{2016ApJ...830..113G} as an expanding bubble formed by an explosion
near the star and carried away by the wind emanating from that star or
a disk around it.

However, in the ''splash'' scenario, we would not only expect material
to be splashing away from the observer, and therefore having less blueshift,
but also some material moving in the plane of the sky, at the same radial velocity as the jet, 
and some even moving faster towards the observer. We do not
observe any emission that is more blueshifted than the core of the jet,
making it difficult to explain the BLVC without resorting to the
ad hoc assumption that the splash was not axisymmetric around the jet.
For this reason, we do not consider this a viable explanation for the
BLVC.

\citet{Machita2008ApJ...676.1088M}
trace the evolution of the density structure, 
the embedded magnetic
field, and the resulting magneto-centrifugal forces from the original
conditions of a small (1 M$_{\sun}$) molecular cloud through its collapse and to the protostellar phase 
with a resistive 3D MHD model
and later refined the models in
\citet{Machita2014ApJ...796L..17M} by running the
model longer to cover the full evolution of the protostellar jet. 
The salient feature of their models is
the formation of both a low-velocity, wide-angle outflow component at densities
of $\approx 1\time10^{11} cm^{-3}$ that they
identify with the source of outflows seen in $CO$ lines emission, and a
much narrower, higher velocity jet component formed at protostellar densities
of $\rm 5\times10^{18} cm^{-3}$ and detected at optical and near-infrared
wavelengths in shock-excited line emission. 
They conclude that the low velocity flow contains a large fraction
of the infalling gas mass and certainly most of the total mass
of the re-ejected outflowing material, as well as most of the angular momentum.

This model bears some resemblance to the two components (jet and BLVC) in our data that
are distinct in spatial location and in velocity. The BLVC 
component of the S(1) emission can be identified with shocks of the low-velocity
outflow component against the possibly still infalling envelope feeding the accretion
onto MHO~3252~Y3. The cross-correlation of the integrated line intensity that was
used to measure the proper motions of the jet knots shows this shock front moving
back towards the star. This is likely caused by changes in brightness of the low
velocity shock, and the superposition with knots in the jet that in combination
appear as a backwards motion in the cross-correlation box. We do not take this
''proper motion'' vector at face value, but rather as an indication that this 
shock is indeed not part of the jet. 

We are left with two possible explanations for the low velocity wider component:
1. This could be a slow moving shock front generated by an eruptive event in the
disk that affected only or mostly the slow moving disk wind, but not the faster
jet that is probably generated closer to the star, or
2. The low velocity feature could be a stationary shock of the slow moving
disk wind with a stationary object, possibly the wide outflow cavity responsible
for the ultimate collimation of the slow wind into a molecular outflow. 

Further proper motion studies of the MHO~3252~Y3 microjet at the highest
available angular resolution will be able to determine whether the BLVC
is a stationary shock or moving in the same direction as the microjet with
about half its speed.

\subsection{Rotation}

The BLVC, labelled ''B'' in Figures 6 and 7 has a lower radial
velocity than the microjet, whose individual knots are labelled
"A, C, D ..." in those figures. The displacement of the BLVC relative
to the jet axis establishes an asymmetry in the velocity field.
However, we do not interpret this fairly large asymmetry of the BLVC relative
to the jet axis as evidence for 
systemic rotation
and angular momentum
transport in the BLVC, but rather ascribe it to the randomness of the 
shock interaction of the BLVC with the outflow cavity and the
ambient cloud that has, apparently, led to more shock excitation
on one side of the BLVC than on the other.

It is plausible that rotation of the microjet itself might be
best observed very close to the central star, before turbulent
interaction leads to a more complicated velocity field.
We do indeed observe slight asymmetries in the collimation region very
close to the MHO~3252~Y3 microjet in Fig. 7, immediately above the inserted
continuum image of the star.
However, this regions very close to the central star is also
most affected by noise from the subtraction of the continuum
emission, and minor apparent asymmetries at the jet base should therefore
not be over-interpreted.

We conclude that we could not convincingly detect any rotation
of the microjet in MHO~3252~Y3.
This is not surprising and should not lead to generalized conclusions
about microjet rotation since our proper motion and radial velocity measurement
indicate that the MHO~3252~Y3 microjet is pointing strongly towards the observer
and the radial velocity effect of jet rotation is perspectively reduced.
Consequently, this object is not well suited for a detection of
microjet rotation, even if such such rotation existed.

\section{CONCLUSIONS}
We have presented photometric data on the variability of the outflow
and microjet source MHO~3252~Y3 and
have found that a complex double-minima light curve with a period of 904~d can
explain the existing observations. This phased light curve predicts
that the next deep minimum will be observable in 2019.
We have studied the proper motion of H$_2$ S(1) emission knots in
the microjet and large-scale outflow of MHO~3252~Y3 and distinguish emission knots physically
associated with the MHO~3252~Y3 microjet from other S(1) emission knots originating from different
outflow sources.
Together with radial velocity measurements, the proper motion of the
microjet indicates that this jet is mostly pointed towards the observer
and is only inclined by 19$\fdg$6 from the line of sight.
Detailed velocity resolved integral field spectroscopy of the
microjet and its launch region reveals a complex velocity field.
One component, the BLVC, is much less blueshifted than the rest of the microjet,
is spatially much more broadly distributed, 
is offset from the microjet axis,
and shows an apparent proper
motion essentially opposite to that of the microjet emission knots.
We tentatively interpret this component as either the result of an
eruptive event in the disk that mostly affected the slow wind component, 
but not the jet, or as the interaction of the
outer layers of the slow outflow wind with the walls of the outflow cavity.
The microjet itself is well collimated. 
Its velocity structure
indicates that
the outflow velocity is highest on
the center axis of the microjet, and decreases outward from there
due to turbulent entrainment of the slower wide wind component.
No rotational velocity
component in the microjet itself was observed.
The microjet must be collimated within the first 85 au along its axis
into a narrow opening angle. Our observations could not
spatially resolve the collimation region.

\acknowledgments

Most photometric data on MHO~3252~Y3 were obtained at the IRIS telescope of the
Universit\"atssternwarte Bochum on Cerro Armazones,
which is operated under a cooperative agreement between
the "Astronomisches Institut, Ruhr Universit\"at Bochum", Germany
and the Institute for Astronomy, University of Hawaii, USA.
Construction of the IRIS infrared
camera was supported by the National Science Foundation under grant AST07-04954.
We wish to thank
Michael Ramolla
and
Christian Westhues
for
operating the IRIS telescope for the acquisition of the data used in this paper.
While the UKIRT data for this project were acquired, UKIRT was first operated by the University of
Arizona and later by the University of Hawaii.
The Fraunhofer Telescope is located on 
Mt. Wendelstein in Germany and is operated by the Ludwig-Maximilian University in Munich, Germany.
Construction of its 3kk camera was supported by the 'Excellence Cluster Origin and Structure of the Universe'. 
The OSIRIS data presented herein were obtained at the W. M. Keck Observatory,
which is operated as a scientific partnership among the California Institute of Technology,
the University of California and the National Aeronautics and Space Administration.
The Observatory was made possible by the generous financial support of the W. M. Keck Foundation.
This publication makes use of data products from the Two Micron All Sky Survey, which is
a joint project of the University of Massachusetts and the Infrared Processing and Analysis Center/
California Institute of Technology, funded by the National Aeronautics and Space Administration
and the National Science Foundation.
Data from the Centro Astron\'{o}mico Hispano Alem\'{a}n (CAHA) Archive at CAB (INTA-CSIC) were used in the paper.
The MHO catalog is now hosted by D. Froebrich at the University of Kent, U.K.
This work has made use of data from the European Space Agency (ESA) mission
{\it Gaia}, processed by the {\it Gaia}
Data Processing and Analysis Consortium (DPAC).
Funding for the DPAC
has been provided by national institutions, in particular the institutions
participating in the {\it Gaia} Multilateral Agreement.
The Pan-STARRS1 Surveys (PS1) and the PS1 public science archive have been made possible through contributions by the Institute for Astronomy, the University of Hawaii, the Pan-STARRS Project Office, the Max-Planck Society and its participating institutes, the Max Planck Institute for Astronomy, Heidelberg and the Max Planck Institute for Extraterrestrial Physics, Garching, The Johns Hopkins University, Durham University, the University of Edinburgh, the Queen's University Belfast, the Harvard-Smithsonian Center for Astrophysics, the Las Cumbres Observatory Global Telescope Network Incorporated, the National Central University of Taiwan, the Space Telescope Science Institute, the National Aeronautics and Space Administration under Grant No. NNX08AR22G issued through the Planetary Science Division of the NASA Science Mission Directorate, the National Science Foundation Grant No. AST-1238877, the University of Maryland, Eotvos Lorand University (ELTE), the Los Alamos National Laboratory, and the Gordon and Betty Moore Foundation.
We thank the referee, L. A. Zapata, for helpful comments that led
to the additional discussion of the distance based on Gaia data.

\vspace{5mm}
\facilities{Keck I, IRIS, UKIRT, Fraunhofer Telescope}

\appendix

\section{A Gaia DR2 Distance to Serpens South}

The distance to W40 was precisely measured by 
\citet{Ortiz-Leon2017} to be 436.0$\pm$9.2~pc,
and a plausible line of arguments leads to the adoption of
this distance as the best value for Serpens South.
A new and independent estimate for the distance to Serpens South is possible
based on the recently released {\it Gaia} 
\citep{Gaia-2016A&A...595A...1G} 
DR2 catalog. This catalog
contains the best available parallax measurements for all
moderately bright, optically visible stars.

The Serpens South molecular cloud is essentially an opaque
screen at optical wavelength 
and this is the reason why the Serpens South cluster 
had not been noticed until its discovery in the infrared
by Gutermuth 2008 based on Spitzer Space Telescope data.
In the direction of the Serpens South star forming region the
{\it Gaia} catalog, measured at optical wavelengths,
contains almost exclusively foreground stars in front of the
molecluar cloud.
The embedded YSO MHO3252~Y3 is a single, isolated object in the southern
region of the Serpens South molecular filament. It is undetectable at
the optical wavelengths where {\it Gaia} operates. 
One star seen near the perimeter of the heavily obscured Serpens South cluster has
sufficiently low extinction to just become detectable to {\it Gaia}
and this star provides 
a useful additional constraint on the distance to MHO3252~Y3.
This star is listed as {\it Gaia} source 4270246258814194304,
Pan-STARRS PSO J183007.854-020231.657, and 2MASS J18300785-0202318.`
The star is not detected on the DSS B photographic plates. 
It is only weakly detected in Pan-STARRS g band and not listed
in the catalog at this wavelength. The fixed aperture photometry AB magnitudes
in the other Pan-STARRS bands are r = 21.328, i=17.934, z=15.701, y=14.080.
The 2MASS magnitudes, converted to AB magnitudes for consistencey are 
J$_{AB}$=11.447, 
H$_{AB}$=9.631, 
K$_{sAB}$=9.033. 

A detailed discussion of the extinction and the intrinsic spectrum
of this star is beyond the scope of this discussion of its distance.
Suffice it to say that the steep optical spectrum clearly shows that
the star lies behind a screen of substantial extinction, which we
assume is the Serpens South molecular cloud.
The star could be embedded in the molecular cloud or lie behind it,
so its distance measured by {\it Gaia} establishes an upper limit for the
distance of that molecular cloud.

While this
star is quite bright at short and mid-infrared wavelength,
it does not show a long wavelenth excess and therefore was
not listed as a Class 0, I, or II YSO by Gutermuth 2008. 
A search of the Chandra database showed that 
this star has not been detected at X-ray energies and it
is not listed as a YSO in the X-ray selected list of 
\citep{Winston.2018AJ....155..241W}.
The star is not associated with reflection nebulosity on 
deep UKIRT images, unlike many other of the YSOs in the
Serpens South cluster.
The projected position of the star is near the periphery of
the Serpens South cluster. It position makes a physical connection
with the cluster plausible, but does not prove it.
We do not have any positive confirmation that the star
is a member of the Serpens south cluster or otherwise
embedded in that molecular cloud. 
The {\it Gaia} DR2 distance to this star is
 451 $\pm$ 79 pc
When interpreted as an
upper limit for the distance of the Serpens South cluster,
this measurement is consistent
with the Ortiz-Leon value for the distance to W40 and the assumption
that Serpens South is at that same distance.



\end{document}